\newdimen\Parindent\newdimen\Parskip
\Parindent=\parindent\Parskip=\parskip%save old values
\def\@oddhead{}\def\@evenhead{}
\def\@oddfoot{\rm\rightmark \hfil Page \thepage}
\def\@evenfoot{\rm\leftmark Page \thepage \hfil}

\newdimen\Parindent\newdimen\Parskip
{\Parindent=\parindent\Parskip=\parskip%save old values
\parindent0pt\parskip3mm\columnsep11mm}%set new values
{\parindent\Parindent\parskip\Parskip}%end{Abstracts}
{\Parindent=\parindent\Parskip=\parskip%save old values
\parindent0pt\parskip3mm\columnsep11mm}%set new values
{\parindent\Parindent\parskip\Parskip}%end{Questions}

\gdef\abstract#1{\gdef\@abstract{#1}}

\def\maketitle{\par
 \begingroup
 \def\thefootnote{\fnsymbol{footnote}}
 \def\@makefnmark{\hbox
 to 0pt{$^{\@thefnmark}$\hss}}
 \twocolumn[\@maketitle]
 \@thanks
 \endgroup
 \setcounter{footnote}{0}
 \let\maketitle\relax
 \let\@maketitle\relax
 \gdef\@thanks{}\gdef\@author{}\gdef\@title{}\let\thanks\relax}
\def\@maketitle{\vbox to 1in{\hsize\textwidth
 \linewidth\hsize \vfil \centering
 { \large\bf \@title \par} \vskip 2em {\begin{tabular}[t]{c}\@author
\end{tabular}\par}
 \vfil}
\hsize\textwidth \linewidth\hsize
\begin{center} ABSTRACT \end{center} \par \small\rm \@abstract \vskip 2em }

\def\section{\@startsection {section}{1}{\z@}{-3.5ex plus -1ex minus
 -.2ex}{2.3ex plus .2ex}{\normalsize\bf}}
\def\subsection{\@startsection{subsection}{2}{\z@}{-3.25ex plus -1ex minus
 -.2ex}{1.5ex plus .2ex}{\normalsize\bf}}
\def\subsubsection{\@startsection{subsubsection}{3}{\z@}{-3.25ex plus
-1ex minus -.2ex}{1.5ex plus .2ex}{\normalsize\bf}}
\def\paragraph{\@startsection
 {paragraph}{4}{\z@}{3.25ex plus 1ex minus .2ex}{-1em}{\normalsize\bf}}
\def\subparagraph{\@startsection
 {subparagraph}{4}{\parindent}{3.25ex plus 1ex minus
 .2ex}{-1em}{\normalsize\bf}}

\documentstyle[HEP93]{article}
\renewcommand{\thefootnote}{\fnsymbol{footnote}}
\begin{document}
\title{IMPACT PICTURE MODEL FOR $\bar{p}p$ ELASTIC SCATTERING}
\author{C. BOURRELY\\
        \it Centre de Physique Th\'eorique\\
        \it CNRS-Luminy case 907, 13288 MARSEILLE Cedex 9, FRANCE
        \
        }
\abstract{\rightskip=1.5pc
          \leftskip=1.5pc
The predictions of an impact-picture model for proton-antiproton elastic
scattering are compared with the new UA4 experimental data in the
Coulomb interference region at the center-of-mass energy of $541 GeV$.
The predicted differential cross section, the parameter $\rho$ and the
forward slope are in very good agreement with the data.}

\maketitle
\vskip 5mm
{\bf \noindent }
In the year 1979 an impact-picture model [1] was presented for the
description of $\bar{p}p$ and $pp$ elastic scattering at high energies.
 This model is based on a Chou and Yang [2] hypothesis assuming that a
proportionality should exist between the momentum transfer dependence
of the electromagnetic form factor of the proton and its hadronic matter
distribution. The energy dependence is built   from the rigourous results
obtained by Cheng and Wu [3] from the high energy behavior of massive
QED. An analysis made in 1984 [4] has allowed a determination of the
 6 parameters wich determine the nuclear amplitude.
Since then, the model predicts the correct rise of the total cross
 section
$\sigma_{tot}$, the shape of the differential cross section $d\sigma/dt$,
and the integrated elastic cross section $\sigma_{el}$ in an energy range
from the CERN-ISR up to the FNAL [4].

In view of the recent results obtained by the UA4 collaboration [5] for
$\bar{p}p$ at $\sqrt{s} = 541 GeV$
in the Coulomb interference region, it was interesting to compare the
experimental results with the prediction of the model, for that
purpose a Coulomb amplitude in the Born approximation with a West-Yennie
 phase [6] was added to the nuclear amplitude.
The experiment gives only
the event rate $dN/dt$  proportional to $d\sigma/dt$ up to a
 normalization
constant which was fixed for a best agreement with the theoretical
prediction in the entire $t$-range.
{}From the result  shown in fig. 1,
one can see that the agreement is excellent.
This new experiment gives also the following values for the ratio $\rho$
of the real to imaginary parts of the forward nuclear amplitude and the
slope B of the forward peak, namely:
\begin{equation}
\rho = 0.135 \pm 0.007~~~\mbox{and}~~~B = (15.4 \pm 0.1)GeV^{-2}
\end{equation}
to be compared with the 1984 prediction [4,7]
\begin{equation}
\rho = 0.13~~~\mbox{and}~~~B = 15.6GeV^{-2}
\end{equation}

In conclusion, the impact-picture model proposed 14 years ago has
 survived the successive experiments on $\bar{p}p$ and $pp$ elastic
 scattering,
 the next generation of colliders like LHC and SSC will provide
 crucial tests for the model.
\begin{figure}[ht]
\vskip 0.5cm
\vspace{10cm}
 \includegraphics{epsua3.ps}
\end{figure}

  \vskip 2.0mm
{\bf \noindent References}
\vskip 1.0mm

\noindent
[1] C. Bourrely, J. Soffer and T.T. Wu,
{\it Phys. Rev.} {\bf D19} (1979) 3249.
\newline
\noindent
[2] T.T. Chou and C.N. Yang,
{\it Phys. Rev. Lett.} {\bf 20} (1968) 1213; {\it Phys. Rev.} {\bf 170}
 (1968) 1591; ibid {\bf 175} (1968) 1832.
\newline
\noindent
[3] H. Cheng and T.T. Wu, {\it Phys. Rev. Lett.} {\bf 22} (1969) 666;
 {\it Phys. Rev.} {\bf 182} (1969) 1852, 1868, 1873, 1899.
\newline
\noindent
[4] C. Bourrely, J. Soffer and T.T. Wu,
{\it Nucl. Phys.} {\bf B247} (1984) 15; {\it Z. Phys.} {\bf C37} (1988)
 369
\newline
\noindent
[5] UA4/2 Collaboration, C. Augier et al.,
\newline
CERN-PPE/93-115, submitted to {\it Phys.lett.}
\noindent
\newline
[6] G.B. West and D.R. Yennie,
{\it Phys. Rev.} {\bf 172} (1968) 1413.
\newline
\noindent
[7] C. Bourrely, J. Soffer and T.T. Wu,
{\it Phys. Lett.} {\bf B315} (1993) 195.

\end{document}